\documentclass{article}

\usepackage{maa-monthly}
\theoremstyle{theorem}
\newtheorem{theorem}{Theorem}
\newtheorem{corollary}{Corollary}
\newtheorem{proposition}{Proposition}

\theoremstyle{definition}
\newtheorem*{definition}{Definition}

\begin{document}

\title{Social Balance and the Bernoulli Equation}
\markright{Social Balance}
\author{J. J. P. Veerman}

\maketitle

\noindent
\section*{Abstract}
\begin{small}
Since the 1940s there has been an interest in the question of why social networks
often give rise to two antagonistic factions. Recently a dynamical model of how and why
such a balance might occur was developed. This note provides
an introduction to the notion of social balance and a new (and simplified)
analysis of that model. This new analysis allows us to choose general initial
conditions, as opposed to the symmetric ones previously considered.
We show that for general initial conditions, four factions will evolve instead of two.
We characterize the four factions, and give an idea of their relative sizes.
\end{small}

%\vskip 0.4in\noindent
\section{Introduction.}
\label{chap:one}
\setcounter{figure}{0} \setcounter{equation}{0}

\noindent
Since the 1940s there has been an interest in the question of why finite social networks
often give rise to two antagonistic factions \cite{CH, Harary, Heider}. Motivated by insights
from the field of social psychology (most notably by Heider \cite{Heider}), Harary \cite{Harary} and
Cartwright and Harary \cite{CH} developed a formal graph-theoretical framework for
social balance that was consistent with Heider's ideas. They were able to prove
that a sign symmetric network is balanced if and only if it has (at most) two factions
(see below for the definitions). This became know as the \emph{structure theorem}.
Until recently however, a dynamical model of how and why such a balance occurs was lacking.

The field received an impetus in 2011 when in \cite{Marvel},
Marvel et al. successfully analyzed the dynamics of the matrix differential equation
\begin{equation}
\dot X= X^2  ; \quad X(0)=X_0.
\label{eq:bernoulli}
\end{equation}
The interpretation of this model is that the entries of the $n\times n$ matrix $X$
represent the opinions of individuals $\{1,\ldots , n\}$ towards one another. The
value of the entry $x_{ij}$ of $X$ indicates the \emph{strength of the friendliness}
\cite{Marvel} of individual $i$ towards individual $j$. The modeling becomes clear
if we write out the differential equation for one entry:
\begin{equation}
\dot x_{ij}=\sum_{k=1}^n\, x_{ik}x_{kj}.
\label{eq:bernoulli-component}
\end{equation}
Thus individual $i$ communicates somehow with all individuals $k\in\{1,\ldots , n\}$. If $i$'s feeling
about $k$ is positive, then $k$'s feeling about $j$ will pull $i$'s opinion in the same
direction. On the other hand, if $i$'s opinion about $k$ is negative, then
$i$'s opinion about $j$ will change in the direction opposite to $k$'s opinion about $j$.
This ``implies roughly that ... one's friends' friends will tend to become one's
friends and one's enemies' enemies also one's friends, and one's enemies' friends and
one's friends' enemies will tend to become one's enemies" \cite{Rapo}.
Since friendliness is difficult to quantify, one is led to study the problem with the initial
value $X_0$ being a \emph{random} matrix.

To perform the analysis of equation (\ref{eq:bernoulli}), the authors of \cite{Marvel} assumed
the matrix $X_0$ is symmetric. (This was later extended to \emph{normal} matrices \cite{Patrick}.)
The analysis itself now proceeded in two parts. The first is to solve the differential equation
(\ref{eq:bernoulli}) by considering it as a special case of a \emph{matrix Riccati equation}
(see \cite{Reid}). This solution is subject to certain conditions, the most important
of which are that $X_0$ has a positive real eigenvalue, and that the largest of these
real eigenvalues, $\lambda$,
has algebraic and geometric multiplicity 1. The second part of the analysis is to show that
for a random symmetric matrix $X_0$ these hypotheses hold with probability tending to 1
as the dimension $n$ grows unbounded. This involves relying on fairly subtle
arguments about eigenvalues of random symmetric matrices (see \cite{Arnold}).

In this article we propose a simpler and more precise treatment of this problem.
In Section 2 we give a very simple proof of a generalization of the structure theorem.
In it the hypothesis that the network be sign symmetric has been dropped.
In Section 3 we treat equation (\ref{eq:bernoulli})
not as a Riccati equation but as a special case of a \emph{Bernoulli equation}
(first mentioned in \cite{Ber}, see also \cite{Teschl} and other
contemporary textbooks). This simplifies our treatment and has the advantage that we can
characterize much more precisely than previously what happens for nonsymmetric (or nonnormal)
initial conditions $X_0$. We make use of a recent study (\cite{Tao}) of random
(nonsymmetric) matrices to assert that also in this case with probability
tending to 1 as the dimension $n$ tends to infinity, $X_0$ has a leading real, simple,
positive eigenvalue. As a result, we are able to show that with overwhelming probability
a random initial condition $X_0$ (not symmetric or normal) will lead to four factions.
Two of these factions are similar to the factions in the symmetric case:
they are \emph{cohesive} in the sense that members of the same faction like each other.
The other two factions are not: members of the same faction dislike each other.
Finally, in Section 4, we show that if the two cohesive factions join forces, they
can be expected to have a narrow majority.
This last result is new.

Finally we remark on a later development. In \cite{Patrick} a slightly different model for
social balance is proposed, namely the differential equation in (\ref{eq:bernoulli})
is replaced by $\dot X= XX^T$. This model gives rise to \emph{two factions} with high probability
under general initial conditions. The analysis is substantially more complicated.
It appears that the simplification we propose here does not help in the study of this model.

%\vskip 0.4in\noindent
\section{What is Social Balance?}
\label{chap:balance2}
\setcounter{figure}{0} \setcounter{equation}{0}

\noindent
The \emph{signed directed graph} $G$ on $n$ vertices is a collection $V$ of $n$ vertices
together with a set $E$ of directed edges between certain ordered pairs of vertices whose
weights are positive or negative.
The graph $G$ is \emph{sign symmetric} if for every pair $u$ and $v$ in $V$, all
weights of any edges between $u$ and $v$ are either all positive or all negative.
An \emph{undirected path or cycle} in a directed graph $G$ is a non-self-intersecting path or cycle
following edges without any regard for their direction. A cycle or path is called \emph{positive} if
the product of the weights encountered along the cycle or path is positive, and \emph{negative} if that
product is negative. A \emph{weakly connected component} of $G$ is a maximal subgraph all of whose
vertices can be connected by undirected paths.

\noindent
\begin{definition} A directed graph $G$ is called \emph{balanced} if
every undirected cycle is positive.
\label{def:balance}
\end{definition}

\vskip -.2in
In the context of social dynamics, this definition is essentially due to Harary \cite{Harary} and
Cartwright and Harary \cite{CH} expanding on earlier concepts by Heider \cite{Heider}.
The idea in \cite{CH} is that presumably in a state of imbalance ``pressures will arise to change it
toward a state of balance."

\noindent
\begin{definition} A directed graph $G$ is said to have \emph{two factions} if the vertices can be
partitioned into (at most) two sets $U_0$ and $U_1$ such that all edges entirely within $U_0$
or within $U_1$ have positive weights, while those that connect the two have negative weights.
\label{def:two factions}
\end{definition}

The main result here is the simple but surprising conclusion that these two definitions are
equivalent for undirected graphs. This result, the \emph{structure theorem}, was proved
in the 1950s \cite{CH, Harary} for undirected graphs. We give a slight generalization here.

\noindent
\begin{theorem} A directed graph with signed weights $G$ is balanced if and only if it has
two factions $U_0$ and $U_1$.
\label{thm:balance}
\end{theorem}

\vskip -.1in\noindent
\begin{proof} Without loss of generality we may assume that $G$ has one (weakly connected) component.
By definition, $G$ is balanced if and only if every undirected cycle is positive. But that is true
if and only if for all $u$ and $v$ in $V$ any two undirected paths connecting $u$ and $v$ have the same sign.
In turn, this is equivalent to partitioning the vertices in $V$ as follows.
Start with a vertex $u$ and call its faction $U_0$. For any $v\in V$, $v$ is
in $U_\pi$ where $\pi$ is 0 if the undirected paths connecting $u$ and $v$ are positive, and
$\pi$ is $1$ if they are negative. \end{proof}

\vskip 0.0in\noindent
{\bf Historical remark.} From a graph $G$ with two factions we obtain a \emph{bipartite} graph $H$
by deleting all its positive edges. One thus sees without much trouble that the above result
is in fact equivalent to a much earlier result published by D. K\"{o}nig in 1916 \cite{koenig},
namely that a graph is bipartite if and only if it has no odd cycles.

%\vskip 0.4in\noindent
\section{When does Social Balance Evolve?}
\label{chap:evolution}

\noindent
\begin{definition} An $n\times n$ matrix $X$ is called \emph{typical} if it is a real matrix that satisfies
 all of the following conditions:\\
1) $\det X_0 \neq 0$.\\
2) $X_0$ has a positive real eigenvalue.\\
3) The largest positive real eigenvalue $\lambda$ of $X_0$ is simple.
\label{def:typical}
\end{definition}

Let $M_n$ be an ensemble of real $n\times n$ matrices
whose entries have independent Gaussian distribution with mean zero and variance one.
Tao and Vu proved that if $n$ is even, then matrices in $M_n$ will have some real eigenvalues with
probability tending to 1 as $n$ tends to infinity \cite[Corollary 17]{Tao}. For this reason
we assume from now on that $n$ is even and use the characterization \emph{typical} in the above definition.
In addition, most of these eigenvalues will be simple \cite[Corollary 18]{Tao}.
Tao and Vu also conjecture, that in fact with overwhelming probability \emph{none} of the
eigenvalues should be repeated.

\begin{figure}[pbth]
\begin{center}
\includegraphics[width=.50\linewidth]{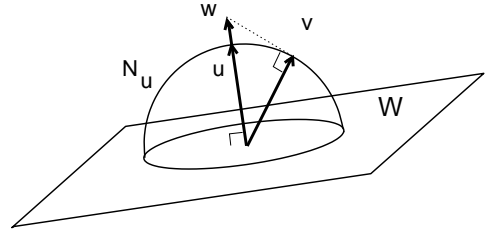}
\caption{ \emph{ Schematic representation of $W$, $N_u$, $w$, $u$, and $v$ in $\mathbb{R}^n$. }}
\label{fig:evecs}
\end{center}
\end{figure}

To facilitate the proof of the next results, we introduce some notation (see Figure \ref{fig:evecs}).
Let $X_0$ be a typical $n\times n$ matrix with leading (real, simple, positive) eigenvalue $\lambda$.
Pick $v$, a unit eigenvector associated with $\lambda$. Denote by $W$ the span of all
(generalized) eigenspaces other than span$\{v\}$. Let $w$ be the vector determined by
\begin{displaymath}
w\in W^\perp \quad {\;\;{\rm and }\;\;} \quad (w,v)=1
\end{displaymath}
where $(\:,\:)$ is the standard inner product in $\mathbb{R}^n$. It will become clear that in fact $w$ is a
\emph{left} eigenvector of $X_0$.
Generally $w$ is not a unit vector. The unit vector in the direction of $w$ will be called $u$.
If $w$ is a unit vector, then we have $w=v=u$. The hyperplane $W$ separates the unit sphere $S^{n-1}$
into two hemispheres, one of which contains the vector $u$.
By construction $u$ and $v$ lie in the same hemisphere which we denote by $N_u$.
Finally we choose an \emph{orthonormal} basis $\{w_2,\ldots , w_n\}$ for $W$.

\begin{theorem}
Let $n$ be even and let $X_0\in M_n$ be a typical $n\times n$ matrix (see Definition \ref{def:typical}).
Then with probability tending to 1 as $n$ tends to $\infty$, the (matrix-valued) initial
value problem
\begin{equation}
\dot X = X^2 ; \quad X(0)=X_0
\label{eq:quadr-model}
\end{equation}
has a unique solution for $t\in[0,\lambda^{-1})$, and near $t=\lambda^{-1}$ the solution
diverges as follows ($v$ and $w$ as defined above):
\begin{equation}
\lim_{t\nearrow \lambda^{-1}} \left(\lambda^{-1}-t\right)\,X = vw^T.
\label{eq:limit}
\end{equation}
\label{thm:main}
\end{theorem}

\vskip -.3in\noindent
\begin{proof} Upon substitution of $Z=X^{-1}$, the differential equation given in
(\ref{eq:quadr-model}) transforms into
\begin{displaymath}
\dot Z = -I ;\quad Z(0)=X_0^{-1}
\end{displaymath}
(where we used condition 1). The solution is:
\begin{equation}
Z(t)=X_0^{-1}-tI \quad {\;\;{\rm or }\;\;} \quad X(t)= \left(X_0^{-1}-tI\right)^{-1}.
\label{eq:soln}
\end{equation}
It is unique and exists until it diverges.

Now recall the notation in the paragraph prior to the statement of Theorem \ref{thm:main}.
Let $P$ be the matrix whose first column is $v$ and whose $i$th column is $w_i$ for $i>1$:
\begin{displaymath}
P=\begin{bmatrix}
v & w_2 & \cdots & w_n
\end{bmatrix} .
\end{displaymath}
By its definition, $P$ is invertible and using conditions 2 and 3 we have
\begin{equation}
X_0=P\,\left( \begin{array}{c|c} \lambda&0^T\\ \hline 0& \tilde X_0 \end{array}\right)\,P^{-1}.
\label{eq:mono-diagonalization}
\end{equation}
This uncouples the first component from the rest of the system. By using equation (\ref{eq:soln}),
we obtain that, for $t\in[0,\lambda^{-1})$,
\begin{equation}
\begin{array}{ccl}
X(t) &=& P\,\left( \begin{array}{c|c} \left(\lambda^{-1}-t\right)^{-1}&0^T\\
\hline 0& \left(\tilde X_0^{-1}-tI\right)^{-1} \end{array}\right)\,P^{-1}\\
 &=& P\,\left( \begin{array}{c|c} \left(\lambda^{-1}-t\right)^{-1}&0^T\\
\hline 0& {\bf 0} \end{array}\right)\,P^{-1} +
P\,\left( \begin{array}{c|c} 0 &0^T\\
\hline 0& \left(\tilde X_0^{-1}-tI\right)^{-1} \end{array}\right)\,P^{-1} \;.
\end{array}
\label{eq:X(t)}
\end{equation}
By hypothesis if $\tilde X_0$ has real eigenvalues, the largest of these is less than $\lambda$, and
so in the interval $[0,\lambda^{-1}]$ the second term is uniformly bounded by a constant.

Finally, $P^{-1}P=1$ implies that the first row $r$ of $P^{-1}$ satisfies
\begin{displaymath}
(r,v)=1 \quad {\;\;{\rm and }\;\;} \quad \rm\;for\; all\; i>1,\; (r,w_i)=0.
\end{displaymath}
Therefore, the first row of $P^{-1}$ equals $w^T$. Equation (\ref{eq:X(t)}) now gives
\begin{displaymath}
X(t)=\left(\lambda^{-1}-t\right)^{-1}vw^T + {\cal O}(1).
\end{displaymath}
Multiplying both sides by $\left(\lambda^{-1}-t\right)$ implies the theorem.
\end{proof}

\vskip 0.1in\noindent
{\bf Remark.} By transposing equation (\ref{eq:mono-diagonalization}), we see that in fact $w$ is a
eigenvector of $X_0^T$ associated with $\lambda$. Equivalently, it is a \emph{left} eigenvector
of $X_0$.

\vskip 0.1in\noindent
{\bf Remark.} The blow-up at finite time may be avoided by dividing the vector field in equation
(\ref{eq:quadr-model}) by $1+|X|$, where $|X|$ is any suitable matrix norm.
The vector fields defined by $X^2$ and $\frac{X^2}{1+|X|}$
have the same direction, and so the phase portrait of the two systems is the same
\cite[Section 1.5]{Chicone}.
The effect of the scalar factor is simply to slow down the flow. Doing this explicitly leads to very
complicated equations without adding insight. We follow the literature on the subject and do not
pursue this.

\begin{corollary} If $X_0$ satisfies the hypotheses of Theorem \ref{thm:main}, then
\begin{displaymath}
\lim_{t\nearrow \lambda^{-1}} \left(\lambda^{-1}-t\right)\,X = vv^T
\end{displaymath}
if and only if the left and right eigenvectors associated to $\lambda$ of $X_0$ are the same.
\label{cor:w_1}
\end{corollary}

This easily implies that for $t$ close to $\lambda^{-1}$ and up to permutation of the coordinates,
$X(t)$ has the block form
\begin{displaymath}
\;{\rm sgn}\; X(t)= \left( \begin{array}{c|c} +&-\\ \hline -&+ \end{array}\right).
\end{displaymath}
Using the terminology of the previous section, this means that the individuals have split into
two factions (one possibly being empty). These factions are \emph{cohesive} in the sense that
individuals within a faction like each other.
The cases discussed in the literature are where $X_0$ is symmetric \cite{Marvel} or normal \cite{Patrick}.
In these cases all eigenspaces are orthogonal, and so we easily recover that $v\in W^\perp$.
One does need to prove that in the more restricted setting of symmetric
or normal matrices, the typical case still has overwhelming probability. This is done
in the literature cited.

We can improve Corollary \ref{cor:w_1} somewhat by noticing the following.

\begin{corollary} If $X_0$ satisfies the hypotheses of Theorem \ref{thm:main},
then for $t$ close to $\lambda^{-1}$, $X(t)$ has two factions if and only if the left and
right eigenvectors associated with $\lambda$ fall in the same \emph{orthant}, i.e., their components
have the same signs.
\label{cor:orthant}
\end{corollary}

%\vskip 0.4in\noindent
\section{What if Social Balance Does Not Evolve?}
\label{chap:non-balance}

\noindent
Even the early sociological literature admits that the assumption that the matrix $X$
is symmetric, or even sign-symmetric, is unrealistic \cite{Heider}. It is therefore natural
to inquire what happens in this model if we drop that assumption.

\begin{figure}[pbth]
\begin{center}
\includegraphics[width=2.3in,height=1.8in]{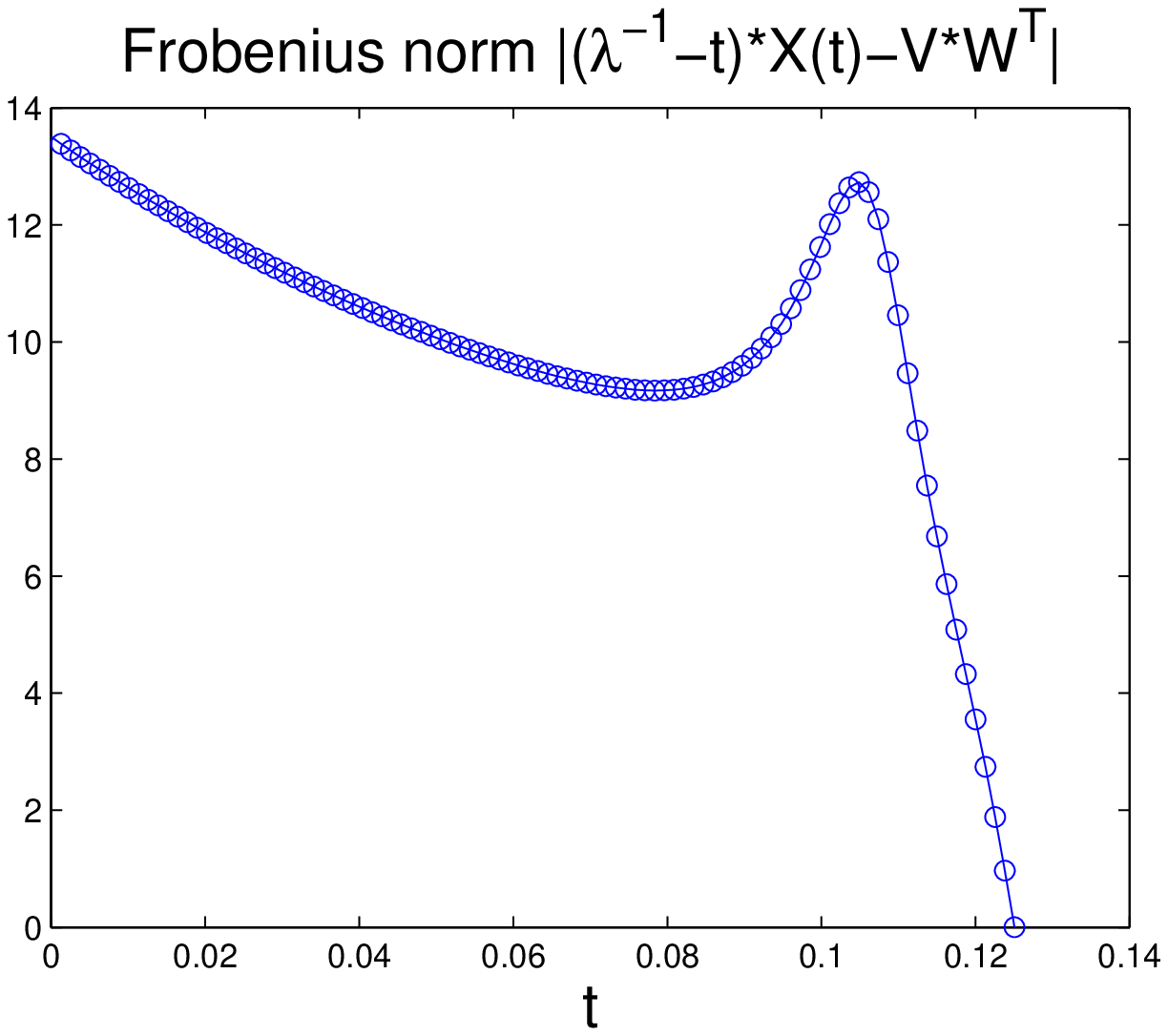}
\includegraphics[width=2.4in,height=1.8in]{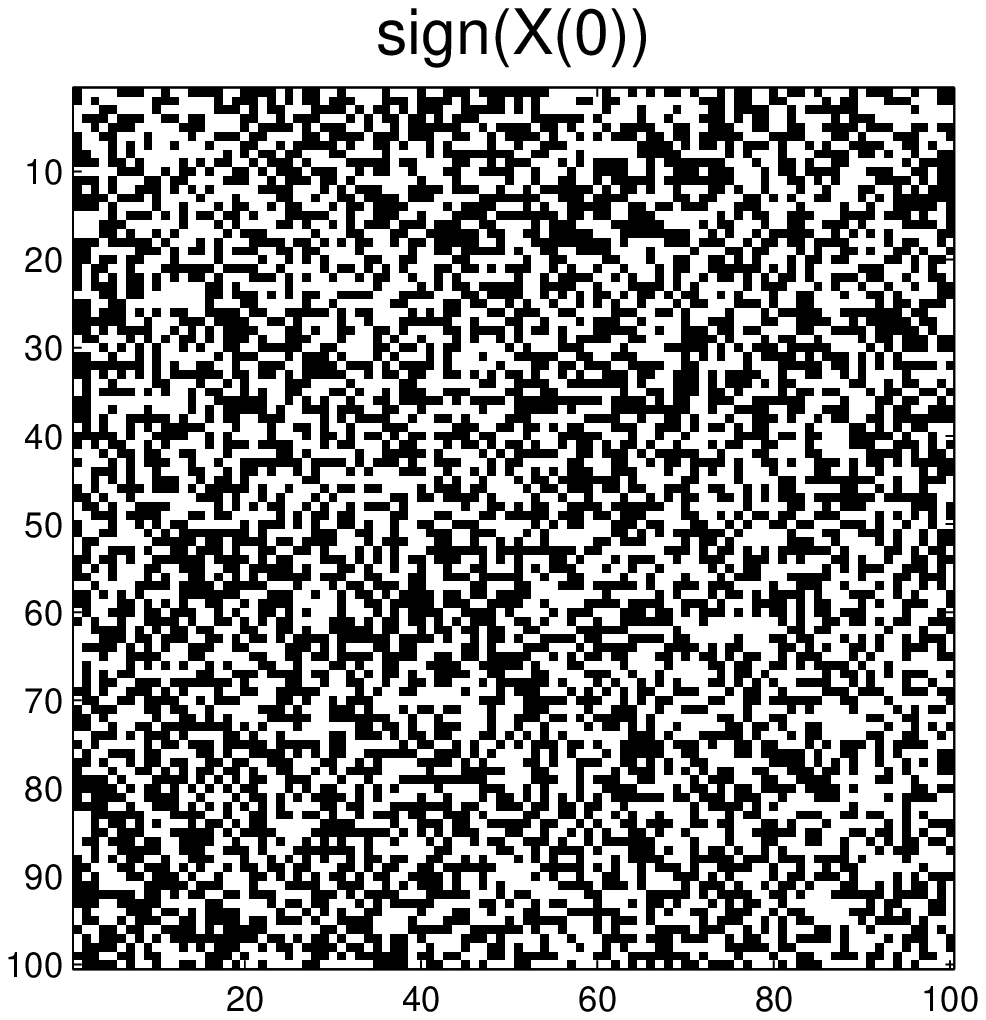}
\includegraphics[width=2.4in,height=1.8in]{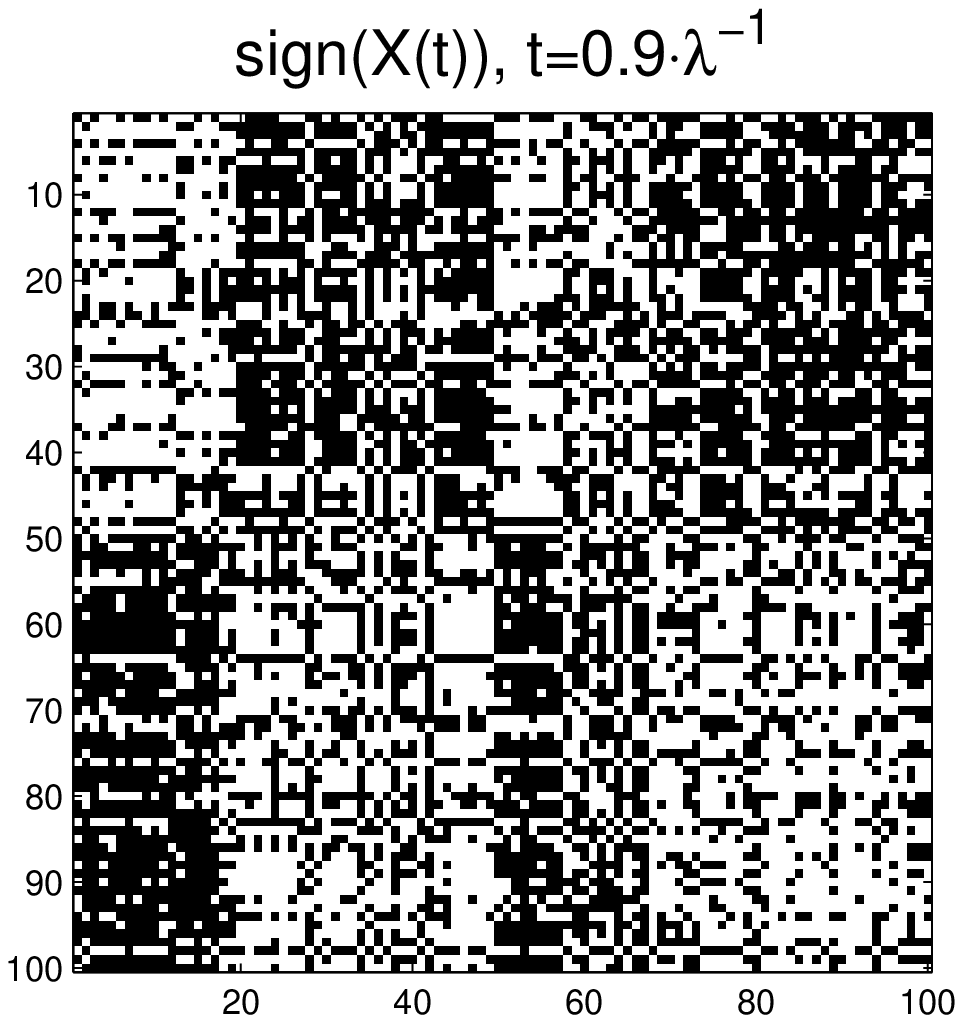}
\includegraphics[width=2.4in,height=1.8in]{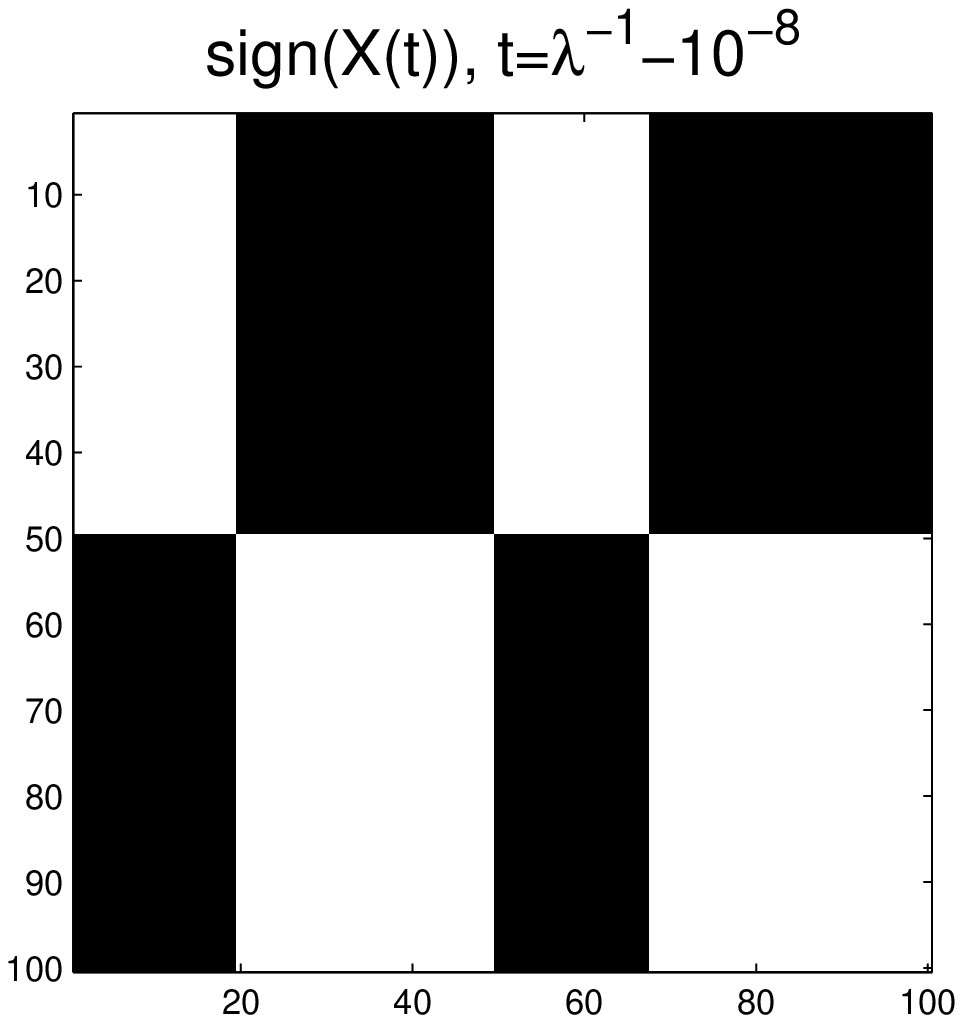}
\caption{ \emph{The upper left diagram illustrates Theorem \ref{thm:main} for a 100 by 100
system ($\lambda\approx 8.01$). The Frobenius norm is the root of the sum of the squares of the
matrix entries. The convergence to the sign pattern of Equation (\ref{eq:final matrix}) is
illustrated in the next three figures. After the appropriate permutation of the components,
we show the signs of the initial configuration $X(t)$ when $t=0$, $t=0.9\cdot \lambda^{-1}$, and
$t=\lambda^{-1}-10^{-8}$, respectively. (White signifies a positive entry, while black signifies negative.)}}
\label{fig:evolution}
\end{center}
\end{figure}

Theorem \ref{thm:main} implies that for $t$ close to $\lambda^{-1}$ the matrix $X(t)$ has the
sign pattern of $vw^T$. Permute the components so that the first $K$
components of $v$ are positive and the others negative. After that, permute the first
$K$ components so that among them, the positive components of $w$ and the negative ones are grouped
together, and so forth. It follows that for $t$ close to $\lambda^{-1}$, $X$ has the
following sign pattern:
\begin{equation}
\;{\rm sgn}\; X(t)= \left( \begin{array}{cc|cc} +&-&+&-\\  +&-&+&-\\\hline -&+&-&+\\  -&+&-&+ \end{array}\right).
\label{eq:final matrix}
\end{equation}
Notice that this matrix is not sign symmetric and thus the system is not balanced. The evolution towards this pattern is illustrated in Figure \ref{fig:evolution}.

Thus there still are two balanced or \emph{cohesive} factions (that is, factions within each of
which individuals like each other). But now there are also two \emph{dispersive} factions
that are internally divided in the sense
that individuals within either one of these factions dislike each other (they even dislike themselves), signifying diagonal blocks of negative entries
(entries $(2,2)$ and $(3,3)$ in equation (\ref{eq:final matrix})).
Note also that the dispersive factions have symmetric feelings about each other
(for example, entries $(2,3)$ and $(3,2)$ are both positive), but the relations
between a cohesive and a dispersive faction are anti-symmetric (for example, entries $(2,4)$ and
$(4,2)$ have opposite signs).

Let us name the factions 1, 2, 3, and 4, in the order of their appearance along the diagonal
of equation (\ref{eq:final matrix}). Adhering to the interpretation that $x_{ij}$ corresponds to
the opinion of $i$ towards $j$, one might speculate that the fact that faction 2 is disliked
by faction 1 while \emph{they} still have positive feelings for faction 1, perhaps caused them
to feel ill at ease and have a negative opinion about themselves. At the same time, they cannot join
faction 4 because they dislike them. Mutual sympathy exists, however, between factions 3 and 4.
Whether or not this is reasonable from a sociological point of view, is unclear to us.

This brings us to the interesting question of which alliances of two factions out of four would
likely give a majority? By symmetry, eigenvectors $v$ and $w$ are equally likely
as $-v$ and $-w$. This would swap factions 1 and 4 on the one hand, and factions 2 and 3 on the other.
Thus the cohesive factions 1 and 4 have the same expected size, and the same holds for the
dispersive factions. The alliance of factions 1 and 2 has the same size as the number
of ``+" signs in the vector $v$. Again by symmetry, the expectation is exactly one half times $n$.
Therefore they will form a majority half the time.
Therefore the same is true for any combination of a dispersive and a cohesive faction.
However, if the \emph{larger of the two cohesive factions} combines with any of the dispersive
factions, then the expectation is that it will have a majority.

The hard part is to decide what will happen if the dispersive factions (factions 3 and 4) join forces,
as was suggested above.
In the remainder of the section, we argue that if the two dispersive factions get together, they are still expected to be a minority.

We use the notation of Section \ref{chap:evolution}
(see also Figure \ref{fig:evecs}).

\begin{theorem} Let $X_0$ satisfy the hypotheses of Theorem \ref{thm:main}. Assume furthermore
that $v$ is uniformly distributed in $N_{u}$. Then the expected number of ``+" signs on the
diagonal of the matrix in equation (\ref{eq:final matrix}) lies in the interval
$\left(\frac{n}{2}, \frac{n}{2}\left(1 + \sqrt{\frac{2}{\pi n}}\right)\right)$.
\label{thm:expectation disgreement}
\end{theorem}

\vskip -.1in
Define
$f_u:N_u\rightarrow \mathbb{Z}$ by
\begin{displaymath}
f_u(v)=\sum_{i=1}^{n}\, \;{\rm sgn}\;  v_i \;{\rm sgn}\; w_i =\sum_{i=1}^{n}\, \;{\rm sgn}\;  v_i
 \;{\rm sgn}\; u_i = n-2k,
\end{displaymath}
where $k$ is the number of sign disagreements between the components of the vectors $v$ and $u$.
Recall that by definition $v$ lies in $N_u$ (see Figure \ref{fig:evecs}). If $v$ is uniformly
distributed in $N_{u}$, then the expected value of $n-2k$ equals
\begin{displaymath}
I^+_u=\frac{1}{\;{\rm vol}\;  N_u}\,\int_{N_u}\,f_u(v)\,dS^{n-1},
\end{displaymath}
where $dS^{n-1}$ is the density of the Lebesgue measure on $S^{n-1}$. In the following, $I^-_u$ denotes
the integral of $f_u(v)$ over the \emph{complement} of $N_u$ in $S^{n-1}$.
Theorem \ref{thm:expectation disgreement} counts the expected number $n-k$ of sign agreements and is
thus a direct consequence of the following proposition.

\noindent
\begin{proposition} Let $n$ be even. Given $u\in S^{n-1}$, then
\begin{displaymath}
0< I^+_u  < \sqrt{\frac{2n}{\pi}}.
\end{displaymath}
\label{prop:integral}
\end{proposition}

\vskip -.2in\noindent
\begin{proof} The first inequality follows from the following two claims:
\begin{displaymath}
I^+_u+I^-_u=0 \quad {\;\;{\rm and }\;\;} \quad I^+_u>I^-_u.
\end{displaymath}
The first of these claims follows from the observation that $I^+_u+I^-_u$ is the integral
over all of $S^{n-1}$. Since $f_u(-v)=-f_u(v)$, this must yield zero.
With probability one, all $u_i$ are nonzero. Therefore, for every $i$, any geodesic arc $\gamma(t)$
connecting $-u$ to $u$ must cross the equator $u_i=0$ exactly once. Thus on such an arc
$\;{\rm sign}\;  u_i \;{\rm sign}\;  v_i|_{\gamma(t)}$ is nondecreasing, and increases from $-1$ to 1. The second claim follows.

To get the other inequality, we first partition $S^{n-1}$ into $2^n$ \emph{orthants} $O_\sigma$. For every
$\sigma\in \{-1,+1\}^{n}$,
\begin{displaymath}
O_\sigma=\{v\in S^{n-1}\,|\, (\;{\rm sgn}\;  u_1 \;{\rm sgn}\;  v_1,\;{\rm sgn}\;  u_2 \;{\rm sgn}\;  v_2,\ldots, \;{\rm sgn}\;  u_{n} \;{\rm sgn}\;  v_{n})=\sigma \}.
\end{displaymath}
Now let the set $Q$ be the quotient of the sequences $\{-1,+1\}^{n}$ and multiplication by $-1$.
Because $n$ is even, $Q$ can be parametrized by those binary sequences $s$ that have positive average
plus half of the ones that have zero average. Define
\begin{displaymath}
Z_s=\left(O_s\cup O_{-s}\right)\cap N_u.
\end{displaymath}
The sets $Z_s$ partition $N_u$ into $2^{n-1}$ sets of equal measure (up to measure zero). For all $v$
in the set $Z_s$, $f_u(v)$ equals either $\sum_{i=1}^{n}\, s_i$ or $\sum_{i=1}^{n}\, (-s_i)$.
Because $\sum_{i=1}^{n}\, s_i$ is equal to $n-2k$ where $k$ is the number of sign differences,
\begin{displaymath}
I^+_u\leq 2^{1-n}\, \sum_{s\in Q}\,\left( \sum_{i=1}^{n}\, s_i\right)=
 2^{1-n}\, \sum_{s\in Q}\,\left(n-2k\right)
\end{displaymath}
 in $Z_s$.
Because $k=\frac n2$ ($n$ is even) contributes 0 to the above sum, we can replace the sum over $Q$
by the sum over $Q'$, the set of \emph{all} sequences in $\{0,1\}^n$ whose average is at least zero.
There are $\binom{n}{k}$ distinct sequences $s$ in $Q'$ that have $k$ sign differences, so
\begin{displaymath}
\begin{array}{ccccc}
2^{1-n}\, \sum_{s\in Q'}\,\left(n-2k\right) &=&
2^{1-n}\, \sum_{k=0}^{\frac n2}\, \binom{n}{k}\left[(n-k)-k\right] & &\\[0.1in]
&=& 2^{1-n}\, \sum_{k=0}^{\frac n2}\, n\left[\binom{n-1}{k}-\binom{n-1}{k-1}\right] &=&
2^{-n}\,n\, \binom{n}{\frac n2}
\end{array}
\end{displaymath}
by telescoping (with the convention that $\binom{n-1}{-1}:= 0$). Now use Stirling's formula in the following form \cite[Section 3.6]{MW}:
\begin{displaymath}
n!=\sqrt{2\pi n}\,\left(\frac ne \right)^n\,\left(1 + \frac{1}{12n} + \frac{1}{288n^2}+\cdots\right).
\end{displaymath}
Proposition \ref{prop:integral} follows after some straightforward algebra.
\end{proof}

\vskip 0.2in
The prediction of Theorem \ref{thm:expectation disgreement} depends on the assumption that
$v$ is uniformly distributed in $N_{u}$. If the distribution $d\rho$ of $v$ is more biased towards
the vicinity of the pole $u$ in $N_u$, then the majority of the united cohesive factions will be more pronounced.
On the contrary, if the distribution is more biased towards the equator, then
that majority will be narrower than described here.

%\vskip .1in\noindent
\begin{acknowledgment}{Acknowledgment.} I am grateful to Patrick de Leenheer for his
insightful comments and indebted to Dacian Daescu for providing the software used to make
figure displaying the simulation.
\end{acknowledgment}

%\vskip 0.4 in

\begin{biog}
\item[J. J. P. Veerman] received his Ph.D. from Cornell University. He has held 
visiting positions in the U.S. (Rockefeller University, Georgia Tech, Penn State), 
as well as in Spain, Brazil, Italy, and Greece. He is currently at Portland State 
University in Oregon, USA, where he is Professor of Mathematics.
\begin{affil}
Maseeh Department of Mathematics and Statistics, Portland State University, Portland, OR 97201.\\
veerman@pdx.edu
\end{affil}
\end{biog}

\vspace{\fill}
\end{document}